\documentclass[aps,reprint,showpacs,superscriptaddress,
footinbib,citeautoscript]{revtex4-1}

\usepackage{graphicx}

\usepackage[utf8]{inputenc}
\usepackage{csquotes}
\usepackage[american]{babel}
\usepackage[T1]{fontenc}
\usepackage{enumerate}
\usepackage{mdwlist}
\usepackage[activate=normal]{pdfcprot}
\usepackage{bbding}
\usepackage{color}
\usepackage{amssymb}
\usepackage{amsmath}
\usepackage{amsfonts}
\usepackage{mathrsfs}

\usepackage{bm}
\usepackage{dcolumn}
\usepackage{color}
\usepackage[colorlinks=true,citecolor=blue]{hyperref}
\hypersetup{colorlinks=true,citecolor=blue,linkcolor=red
,urlcolor=blue}

\begin{document}

\title{Thermoelectric effects in topological crystalline insulators}

\author{Babak Zare Rameshti}
\email{b.zare.r@ipm.ir}
\affiliation{School of Physics, Institute for Research in Fundamental Sciences (IPM), Tehran 19395-5531, Iran}
\author{Reza Asgari}
\affiliation{School of Physics, Institute for Research in Fundamental Sciences (IPM), Tehran 19395-5531, Iran}
\affiliation{School of Nano Science, Institute
for Research in Fundamental Sciences (IPM), Tehran 19395-5531,
Iran}

\begin{abstract}
We investigate the electrical conductivity and thermoelectric effects in topological crystalline insulators in the presence of short- and long-range impurity interactions. We employ the generalized Boltzmann formalism for anisotropic Fermi surface systems. The conductivity exhibits a local minimum as doping varies owing to the Van Hove singularity in the density of states originated from the saddle point in the surface states band structure. Suppression of the interband scattering of the charge carriers at high-energy Dirac points results in a maximum in the electrical conductivity. Whenever the Fermi level passes an extremum in the conductivity, Seebeck coefficient changes sign. In addition, it is revealed that profound thermoelectric effects can be attained around these extrema points.

\end{abstract}

\maketitle

\section{Introduction}\label{sec:intro}
Since the discovery of topological insulators (TI) protected by time-reversal symmetry~\cite{TI}, the extension of the topological classification to other discrete symmetry classes such as particle-hole and translation symmetries has been investigated. Recently, this classification has been extended to topologically distinct classes of band structures which cannot be smoothly deformed into each other without breaking certain crystal point group symmetries~\cite{fu}. A hallmark of this new class, topological crystalline insulators (TCIs), is the existence of the surface states on crystal faces which respect the corresponding crystal symmetry~\cite{fu, mong}. The low-energy properties of the surface states are determined by the surface orientation. The Dirac structure of the TCI is quite distinct from that in the standard TI systems. The TCI supports an even number of metallic Dirac states on crystal surfaces, while it is an odd number of the most common TI systems. There are two types of the surface states~\cite{liu2013}, which preserve the discrete rotational symmetries and support gapless surface states, in which type I refers to the (110) surface states and type II denotes the states of (001) and (111) surfaces.
\par
Hsieh \textit{et al} predicted \cite{hsieh} that, using first-principles simulations, the IV-VI semiconductor SnTe as well as related alloys Pb$_{1-x}$Sn$_{x}$Te and Pb$_{1-x}$Sn$_{x}$Se belong to TCIs class protected by mirror symmetry~\cite{teo} which is characterized by robust surface states on the (001) plane. The inverted band ordering in SnTe, in which the valence band is originated from the $p$ orbitals of the cation Sn and the conduction made by the band from Te, relative to a trivial ionic insulator gives rise to the TCIs phase in SnTe~\cite{hsieh}. Remarkably, this prediction has been experimentally confirmed by the direct observation of topological surface states using the angle-resolved photoemission spectroscopy (ARPES)~\cite{tanaka, dziawa, xu}. Signatures of surface states have also been observed in transport and scanning tunneling microscopy (STM) measurements \cite{ando, okada, gyenis}.
\par
\begin{figure}
\includegraphics[width=8.4cm]{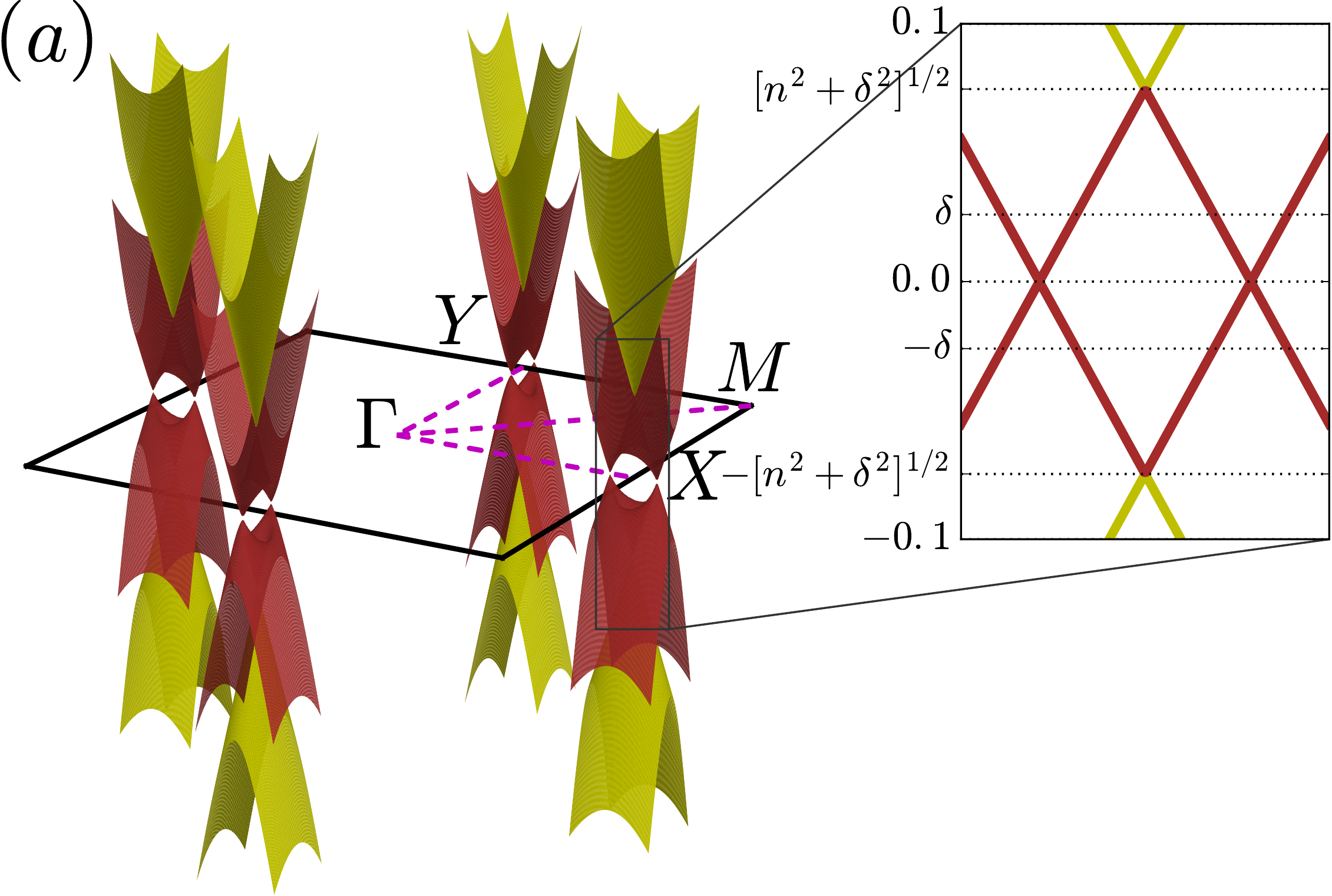}
\includegraphics[width=8.4cm]{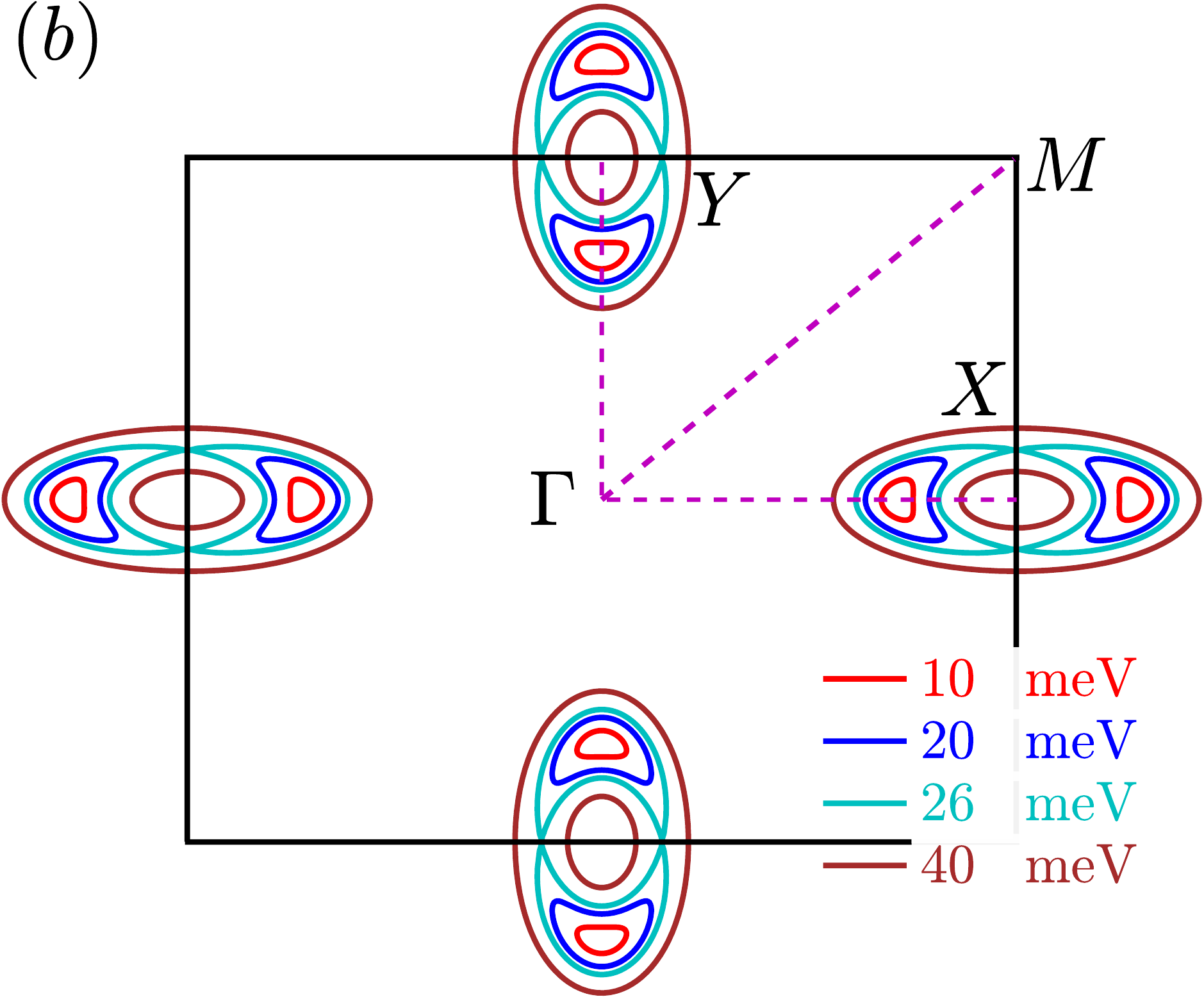}
\caption{(Color online) (a) The TCI (001) surface states band structure in the surface Brillouin zone. The (001) surface states consists of four Dirac cones centered away from the $X$ and $Y$ points, located on the line $\Gamma-X$ and $\Gamma-Y$, respectively. The high-energy Dirac cones are centered at the $X$($Y$) point. Inset shows the detailed band structure in the plane of $\Gamma X$ and in the vicinity of the $X$-point. (b) The corresponding constant-energy contour plots in the surface Brillouin zone. The constant-energy contour undergoes a Lifshitz transition, changing from two disconnected electron pockets to two concentric electron and hole pockets and vice versa, as energy changes away from zero energy.}\label{fig1}
\end{figure}

The remarkable properties of the TCIs is that there emerge four topological protected surface Dirac cones. Alongside with theoretical studies stemming from first-principles simulations and a low-energy effective Dirac theory, it has been verified experimentally \cite{tanaka, dziawa, xu} that in the (001) surface states there are gapless Dirac cones at the $X$ and $Y$ points and at the $\Gamma$ and three $M$ points in the (111) surface~\cite{tanaka2, polley}. The (001) surface states contain four Dirac cones centered at mirror-symmetric-invariant momenta (two along $X-\Gamma-X$ and two along $Y-\Gamma-Y$, as indicated in Fig. \ref{fig1}) in the surface Brillouin zone, while the (111) surface states consist of Dirac cones centered at a time-reversal-invariant momenta, namely; $\Gamma$ and $M$ points. The low-energy Dirac points in the (001) surface centered away from the $X$ and $Y$ points are protected only by mirror symmetry, {\it i.e.}, the two degenerate states at each Dirac point have different mirror eigenvalues, while the high-energy Dirac points centered at the $X$ ($Y$) point are only protected by time-reversal symmetry. As the energy changes away from the low-energy Dirac-point, the topology of the surface states band structures undergoes a Lifshitz transition at a critical energy where the constant-energy contour changes from two separate charge pockets to two concentric pockets. At this transition point, saddle points in the surface band structure lead to a Van Hove singularity in the density of states.
\par
A hallmark of the thermoelectric material study came actually from the pioneering works in Ref.~[\onlinecite{HD}] proposing that nanostructuring materials should most likely provide high thermoelectric efficiencies than with bulk materials. To the best of our knowledge, no theoretical and experimental works have
been carried out on the conductivity and thermoelectric properties of TCIs. Theoretical and numerical studies were mainly focused on the band structure calculations and $k\cdot p$ Hamiltonian for various types of TCIs surface states. Thermoelectric effects have received great attention in recent years owing to their crucial relevance in meso- and nanoscopic systems \cite{giazotto,dubi}. The study of the thermoelectric is not only technologically helpful in managing the generated heat in nano-devices, but also the analysis of thermoelectric effects are of fundamental interest as it is very useful in elucidating some details of the electronic band structure of TCIs surface states that cannot be probed by conductance measurements alone due to a particular to the ambipolar nature of this gapless material.
\par
In this paper, we investigate the combination of the charge and heat transport in TCI (001) surface states as a system due to their peculiar features. We consider the system exposed by a thermal gradient and moreover bias voltages in diffusive regimes considering a short-range potential and long-range charge-charge Coulomb potential with a Thomas-Fermi screening as the source of scattering. Having employed the generalized Boltzmann transport equation, we obtain the conductivity, the Seebeck coefficient and the figure of merit which is a measure of thermoelectric efficiency. Our findings show that when the Fermi energy lies at the transition point, a local minimum of the conductivity emerges and on the other hand, when it lies at the high energy Dirac points, the inter-band scattering is suppressed due to the vanishing density of states for one of the subband at that point. As a result, a maximum at a finite doping appears in the conductivity. In addition, the Seebeck coefficient changes sign at the low- and high-energy Dirac points as well as transition points, owing to the conversion of the charge carriers from electrons to holes and vice versa. Taking into account the fact that electronic properties of TCI surface states can be easily tuned in contrast to common materials due to the gapless excitation spectrum of TCI surface states, these results propose that TCIs could be a promising material for the caloritronic applications rather than common metals.
\par
This paper is organized as follows. In Sec.~\ref{sec:model}, we introduce our system and model and explain the method which is used to calculate the conductivity and thermoelectric coefficients in the presence of short-range and long-range Coulomb potentials using the generalized Boltzmann method. In Sec.~\ref{sec:results}, we present and describe our numerical results for the conductivity and thermoelectric coefficients of TCI surface states. Finally, we conclude and summarize our main results in Sec.~\ref{sec:concl}.

\section{Model and Basic Formalism}\label{sec:model}

\subsection{Hamiltonian}
Topological crystalline insulators are characterized by robust surface states on the $(001)$ plane, which in the surface Brillouin zone contains four Dirac points centered on non-time reversal invariant momenta at the low-energy, two of them are along the $X-\Gamma-X$ direction and two others are along the $Y-\Gamma-Y$ direction. The minimal Hamiltonian near the $X$ and $Y$ points to capture all the essential features of the $(001)$ surface states  are given by~\cite{liu2013, fang2013, liu2014, wang2013, ezawa2014},
\begin{eqnarray}
{\cal H}_{X}({\bf k})&=&v_{1}k_{x}\sigma_{y}-v_{2}k_{y}\sigma_{x}+n\tau_{x}+\delta\sigma_{y}\tau_{y}\nonumber\\
{\cal H}_{Y}({\bf k})&=&v_{2}k_{x}\sigma_{y}-v_{1}k_{y}\sigma_{x}+n\tau_{x}+\delta\sigma_{x}\tau_{y}\qquad\label{hxy}
\end{eqnarray}
Here $\sigma$ and $\tau$ are the Pauli matrices for the spin and pseudospin representing the cation-anion degree of freedom respectively, and the off-diagonal terms $n$ and $\delta$ describe the pseudospin mixing. Typical values are $v_{1}=1.3$~eV$\AA$, $v_{2}=2.4$~eV$\AA$, $n=70$~meV, and $\delta=26$~meV. In the absence of symmetry-breaking perturbations, the $X$ and $Y$ points are related to each other by a rotation of $\pi/2$ since the band structure near the $X$ has a symmetry-related partner near the $Y$ point. This Hamiltonian respects all the symmetries of the $(001)$ surface, including mirror symmetry about the $xz$ plane ($M_{xz}$), mirror symmetry about the $yz$ plane ($M_{yz}$), and time reversal symmetry ($\Theta=\mathcal{T}\mathcal{C}$ where $\mathcal{C}$ denotes a complex conjugate). After diagonalizing ${\cal H}_{X}$ from Eq. (\ref{hxy}), we obtain four surface bands with energy-momentum dispersions in the vicinity of the $X$ point and they are given by
\begin{eqnarray}
\varepsilon_{X}^{2}({\bf k})&=&n^{2}+\delta^{2}+v_{1}^{2}k_{x}^{2}+v_{2}^{2}k_{y}^{2}\nonumber\\
&&\pm2\sqrt{\left(n^{2}+\delta^{2}\right)v_{1}^{2}k_{x}^{2}+n^{2}v_{2}^{2}k_{y}^{2}}\label{ham}
\end{eqnarray}

Furthermore, in the vicinity of the $Y$ point, the energy can be obtained by substituting, $k_{x}\mapsto k_{y},~ k_{y}\mapsto k_{x}$ in Eq. (\ref{ham}). The corresponding surface band structure is shown in Fig. (\ref{fig1}). In the case that $\delta=0$, the lower Dirac cone associated with upper Dirac point overlaps with the upper Dirac cone associated with a lower Dirac point on an ellipsoid in $k$ space at zero energy given by $v_{1}^{2}k_{x}^{2}+v_{2}^{2}k_{y}^{2}=n^{2}$. A nonzero $\delta$ lifts this degeneracy everywhere except for two points along the $x$ axis where two bands with opposite mirror eigenvalues (associated with the reflection $M_{xz}$) cross each other. At low-energy regime, there are two Dirac cones along the $\Gamma-X-\Gamma$ direction located at $(k_{x}, k_{y})=(\pm\Lambda, 0)$ where $\Lambda=\sqrt{n^{2}+\delta^{2}}/v_{1}$, which are protected only by mirror symmetry ($M_{xz}$). However, along the $M-X-M$ direction, there is no band crossing point due to the absence of the mirror symmetry. There is an anti-crossing in that line owing to the fact that they have the same eigenvalues of $M_{yz}$. Two high-energy Dirac points at energies $\varepsilon=\pm\sqrt{n^{2}+\delta^{2}}$, centered at the $X$ ($Y$) point are protected only by the time reversal symmetry and therefore they appear in both directions, as indicated in Fig. (\ref{fig1}).
\par
By invoking the band energy dispersion, the $x$- and $y$- components of the velocity in the vicinity of the $X$ point are given by
\begin{eqnarray}
v_{x}^{X}({\bf k})&=&\frac{v_{1}^{2}k_{x}}{\varepsilon({\bf k})}\left[1\pm\frac{n^{2}+\delta^{2}}{\sqrt{\left(n^{2}+\delta^{2}\right)v_{1}^{2}k_{x}^{2}+n^{2}v_{2}^{2}k_{y}^{2}}}\right],\\
v_{y}^{X}({\bf k})&=&\frac{v_{2}^{2}k_{y}}{\varepsilon({\bf k})}\left[1\pm\frac{n^{2}}{\sqrt{\left(n^{2}+\delta^{2}\right)v_{1}^{2}k_{x}^{2}+n^{2}v_{2}^{2}k_{y}^{2}}}\right].
\end{eqnarray}
and in the vicinity of the $Y$ point those velocities can be obtained by using the following transformation: $v_{x}^{Y}({\bf k})=v_{y}^{X}(k_{y}\mapsto k_{x})$ and $v_{y}^{Y}({\bf k})=v_{x}^{X}(k_{x}\mapsto k_{y})$, due to the underlying fourfold rotation symmetry.

\subsection{Generalized Boltzmann Equations}

In this Section, we consider the transport properties of the conduction electrons of TCIs experiencing thermoelectric forces, a temperature gradient $\nabla T$, an external electric field $E$, and a density gradient $\nabla n$ where the two latest can be combined with an effective field $\bm{\mathcal{E}}=E+\frac{1}{e}\frac{\partial\mu}{\partial n}\nabla n$ where $\mu$ indicates the chemical potential. In the linear response regime where the relation between driving forces and the resulting generalized currents are linear, the response matrix,
\begin{eqnarray}
\begin{pmatrix}
j\\j^{q}
\end{pmatrix}=\begin{pmatrix}
\sigma & \sigma\mathcal{S}\\ T\sigma\mathcal{S} & \mathcal{K}
\end{pmatrix}\begin{pmatrix}
\bm{\mathcal{E}}\\-\nabla T
\end{pmatrix}
\end{eqnarray}
relates the charge $j$ and heat $j^{q}$ currents to the effective electric field $\mathcal{E}$ and the temperature gradient $\nabla T$. Here, $\sigma$ and $\mathcal{K}$ are the electrical and thermal conductivities, respectively, and $\mathcal{S}$ denotes the thermopower which describes the voltage generation owing to the temperature gradient. The two off-diagonal thermoelectric coefficients are related to each other through the Onsager relation.
\par
The figure of merit, the ability of a material to efficiently produce thermoelectric power, is described by a dimensionless quantity denoted by $\mathcal{Z}T$ as
\begin{equation}
\mathcal{Z}T=\frac{\sigma\mathcal{S}^{2}}{\mathcal{K}}T.
\end{equation}
For the practical viability of thermoelectric applications, tailoring materials with high $\mathcal{Z}T$ is certainly the main issue. In this purpose, it is needed the power factor $\sigma\mathcal{S}^{2}$ to be increased at fixed $\mathcal{K}$ value. We explore this factor in detail in TCIs system. Moreover, since the thermal conductivity incorporates both the electron and phonon contributions, we just concentrate on the low enough temperatures where only electrons contribute effectively in the thermal transport and disregard phonon contributions. Though, the phonon thermal conductivity is significantly reduces to interface effects at finite temperature.
\par
Below, we derive general expressions for the charge and heat conductances and thermopower, in the diffusive transport regime, employing the generalized Boltzmann formalism for an anisotropic two-band system~\cite{sinova, faridi}. The thermoelectric properties due to the presence of both electric field and temperature gradient can be found by taking into account two important cases of the short-range (SR) potential (\textit{e.g.}, defects or neutral adatoms) giving by a Dirac delta-like potential and long-range (LR) Coulomb potential in our study. In the diffusive regime, the transport coefficients can be calculated from the following general expression for the charge current and energy flux density,
\begin{eqnarray}
\left[\begin{array}{c}
{\bf j} \\
{\bf j}^{q}
\end{array} \right]=\sum_{n}\int\frac{d^{2}k}{(2\pi)^{2}}\left[\begin{array}{c}
-e \\
\varepsilon_{n}({\bf k})-\mu
\end{array} \right]{\bf v}_{n}({\bf k})f_{n}({\bf k})
\end{eqnarray}
where ${\bf v}_{n}({\bf k})$ is the semiclassical velocity of the carriers in the band $n$ which is related to the energy dispersion $\varepsilon_{{\bf k}, n}$ through ${\bf v}_{n}=(1/\hbar)\nabla_{{\bf k}}\varepsilon_{{\bf k}, n}$. The nonequilibrium distribution function $f_{n}({\bf k})$ describes the evolution of the charge distribution in the presence of an external perturbation. In order to calculate the current densities, we do need to obtain the nonequilibrium distribution function $f=f_{n}({\bf k}, \bm{\mathcal{E}}, T)$ in the presence of driving fields. To this end, we take the Boltzmann equation up to a linear order in the presence of driving fields,
\begin{eqnarray}
&&\hspace*{-10pt}\left(-e\bm{\mathcal{E}}+\frac{\varepsilon-\mu}{T}\nabla T\right)\cdot{\bf v}_{n}({\bf k})\left[-\partial_{\varepsilon}f^{0}(\varepsilon_{{\bf k}_{n}})\right]=\nonumber\\
&&\sum_{n^{\prime}}\int\frac{d^{2}k^{\prime}}{(2\pi)^{2}}w_{nn^{\prime}}({\bf k}, {\bf k^{\prime}})\left[f_{n}({\bf k}, \bm{\mathcal{E}}, T)-f_{n^{\prime}}({\bf k}^{\prime}, \bm{\mathcal{E}}, T)\right]
\label{2dani}
\end{eqnarray}
where $f^{0}$ is the equilibrium distribution function and $w_{nn^{\prime}}({\bf k}, {\bf k^{\prime}})$ is the scattering rate from state ${\bf k}$ in band $n$ to final state ${\bf k}^{\prime}$ in band $n^{\prime}$ which needs to be specified according to the microscopic origin of the scattering mechanisms. The assumption of the elastic scattering, $w_{nn^{\prime}}({\bf k}, {\bf k^{\prime}})\propto\delta(\varepsilon_{{\bf k}, n}-\varepsilon_{{\bf k}^{\prime}, n^{\prime}})$ and microreversibility condition implies that $w_{nn^{\prime}}({\bf k}, {\bf k^{\prime}})=w_{n^{\prime}n}({\bf k^{\prime}}, {\bf k})$. Although, the relaxation time approximation leads to the exact solution of the Boltzmann equation in isotropic systems, this approximative approach cannot describe the full aspects of the anisotropy features of the transport properties. To prevail this, an exact integral equation approach might be implemented. Having parameterized $\bm{\mathcal{E}}$ and ${\bf k}$ as $\bm{\mathcal{E}}=\mathcal{E}(\cos\theta, \sin\theta)$ and ${\bf k}=k(\cos\phi, \sin\phi)$, in the linear response theory, we seek a solution of Eq. (\ref{2dani}) in the form of
\begin{eqnarray}
f_{n}(\phi, \theta)-f^{0}_{n}&=&\left[A_{n}(\phi)\cos\theta+B_{n}(\phi)\sin\theta\right]\mathcal{E}\nonumber\\
&&+\left[C_{n}(\phi)\cos\theta+D_{n}(\phi)\sin\theta\right]\nabla T\label{noneqdis}
\end{eqnarray}
where, $A_{n}(\phi)=\partial_{\mathcal{E}_{x}}f_{n}$, $B_{n}(\phi)=\partial_{\mathcal{E}_{y}}f_{n}$, $C_{n}(\phi)=\partial_{\nabla T_{x}}f_{n}$, and $D_{n}(\phi)=\partial_{\nabla T_{y}}f_{n}$. In the anisotropic bands, ${\bf v}_{n}$ do not need to be parallel with ${\bf k}$, therefore, we use $\xi_{n}(\phi)$ defined by ${\bf v}_{n}({\bf k})=v_{n}(\phi)(\cos\xi_{n}(\phi), \sin\xi_{n}(\phi))$ to parameterize the Fermi velocities of the two bands. By plucking Eq. (\ref{noneqdis}) into Eq. (\ref{2dani}), we end up with the following set of linear integral equations \cite{sinova, faridi}
\begin{eqnarray}
\cos\xi_{n}(\phi)&=&\bar{w}_{n}(\phi)a_{n}(\phi)\nonumber\\
&&-\sum_{n^{\prime}}\int d\phi^{\prime}\frac{v_{n^{\prime}}(\phi^{\prime})}{v_{n}(\phi)}w_{nn^{\prime}}(\phi, \phi^{\prime})a_{n^{\prime}}(\phi^{\prime})\qquad\label{fred1}\\
\sin\xi_{n}(\phi)&=&\bar{w}_{n}(\phi)b_{n}(\phi)\nonumber\\
&&-\sum_{n^{\prime}}\int d\phi^{\prime}\frac{v_{n^{\prime}}(\phi^{\prime})}{v_{n}(\phi)}w_{nn^{\prime}}(\phi, \phi^{\prime})b_{n^{\prime}}(\phi^{\prime})\label{fred2}
\end{eqnarray}
with similar relations for $c_{n}(\phi)$ and $d_{n}(\phi)$. Here $w_{nn^{\prime}}(\phi, \phi^{\prime})=(2\pi)^{-2}\int k^{\prime}dk^{\prime}w_{nn^{\prime}}(k, k^{\prime})$ and $\bar{w}_{n}(\phi)=\sum_{n^{\prime}}\int d\phi^{\prime}w_{nn^{\prime}}(\phi, \phi^{\prime})$. Also, the quantities $A_{n}(\phi)=-ev_{n}(\phi)\left[-\partial_{\varepsilon}f_{n}^{0}\right]a_{n}(\phi)$, $B_{n}(\phi)=-ev_{n}(\phi)\left[-\partial_{\varepsilon}f_{n}^{0}\right]b_{n}(\phi)$, $C_{n}(\phi)=v_{n}(\phi)\left(\frac{\varepsilon-\mu}{T}\right)\left[-\partial_{\varepsilon}f_{n}^{0}\right]c_{n}(\phi)$, and $D_{n}(\phi)=v_{n}(\phi)\left(\frac{\varepsilon-\mu}{T}\right)\left[-\partial_{\varepsilon}f_{n}^{0}\right]d_{n}(\phi)$ are defined. The scattering rates using the Fermi golden rule within the lowest order of the Born approximation is given by
\begin{eqnarray}
w_{nn^{\prime}}({\bf k}, {\bf k}^{\prime})=\frac{2\pi}{\hbar}n_{i}\big\vert\langle n^{\prime}{\bf k}^{\prime}\vert\hat{V}\vert n{\bf k}\rangle\big\vert^{2}\delta(\varepsilon_{{\bf k}, n}-\varepsilon_{{\bf k}^{\prime}, n^{\prime}})\quad
\end{eqnarray}
where $n_{i}$ is the areal density of randomly distributed scatterers and $V({\bf k}-{\bf k'})$
is the Fourier transformation of the interaction potential between an electron and a single impurity, and obviously it depends on the nature of impurities. In the case of short-
ranged it is quite usual to approximate it with a zero-range hard-core potential $V({\bf k}-{\bf k'}) = V_0$. On the other hand, scattering from charged impurities will be long-
ranged Coulombic interaction. This potential is screened by other electrons of the system like Thomas-Fermi approach.
The scattering rates $w_{nn^{\prime}}(\phi, \phi^{\prime})$ thus given by
\begin{eqnarray}
w_{nn^{\prime}}(\phi, \phi^{\prime})&=&\int_{0}^{\infty}\frac{k^{\prime}dk^{\prime}}{(2\pi)^{2}}w_{nn^{\prime}}({\bf k}, {\bf k^{\prime}})\nonumber\\
&=&\frac{n_{i}}{2\pi\hbar}\big\vert\langle n^{\prime}{\bf k}^{\prime}\vert\hat{V}\vert n{\bf k}\rangle\big\vert^{2}\big\vert \nabla_{{\bf k}}\varepsilon_{n}({\bf k})\cdot{\bf k}/k^{2}\big\vert^{-1}.~~\quad
\end{eqnarray}
which is only a function of $\phi$ and $\phi^{\prime}$. By invoking the relaxation-rate-like quantities, solutions of Eqs. (\ref{fred1},\ref{fred2}) into Eq. (\ref{noneqdis}) yield the exact solution of the Boltzmann equation up to the linear order in $\mathcal{E}$ and $\nabla T$.
\par
The response matrix coefficients can be expressed in terms of some kinetic coefficients $\mathcal{L}^{\beta}=\sum_{n}\mathcal{L}_{n}^{\beta}$ as the following,
\begin{eqnarray}
\sigma&=&\mathcal{L}^{0},\quad\mathcal{S}=-\frac{1}{eT}\left(\mathcal{L}^{0}\right)^{-1}\cdot\mathcal{L}^{1},\nonumber\\
\mathcal{K}&=&\frac{1}{e^{2}T}\left(\mathcal{L}^{2}-\mathcal{L}^{1}\cdot\left(\mathcal{L}^{0}\right)^{-1}\cdot\mathcal{L}^{1}\right).\label{thermcoeff}
\end{eqnarray}
All of the coefficients obey the relation
\begin{eqnarray}
\mathcal{L}_{n}^{\beta}\left(\theta, \theta^{\prime}\right)=\int_{-\infty}^{\infty}d\varepsilon\left[-\partial_{\varepsilon}f^{0}_{n}\right]\left(\varepsilon_{n}-\mu\right)^{\beta}\sigma_{n}\left(\varepsilon; \theta, \theta^{\prime}\right)\qquad\label{lls}
\end{eqnarray}
with conductivity given by
\begin{eqnarray}
\hspace*{-20pt}\sigma_{n}\left(\varepsilon; \theta, \theta^{\prime}\right)&=&e^{2}\int\frac{d^{2}k}{(2\pi)^{2}}\delta\left(\varepsilon_{n}-\varepsilon({\bf k})\right)v_{n}^{2}(\phi)\nonumber\\
&&\hspace*{-20pt}\left[a_{n}(\phi)\cos\theta^{\prime}+b_{n}(\phi)\sin\theta^{\prime}\right]\cos\left(\theta-\xi_{n}(\phi)\right)
\end{eqnarray}
with $\theta=\theta^{\prime}=0$ for $\sigma_{xx}$ and $\theta=\theta^{\prime}=\pi/2$ for $\sigma_{yy}$, i.e., longitudinal currents. 

Transverse currents, however, can be derived by taking into account the Berry phase of Bloch states, which has been established in~\cite{dixiao} that has a significant effect on transport driven by thermoelectric forces. The presence of the Berry phase introduces anomalous transport, {\it i.e.}, transport in the transverse direction of a thermoelectric force. In anomalous transport case at low temperatures, we thus have~\cite{dixiao},
\begin{eqnarray}
&&\sigma_{yx}(\varepsilon)=\frac{e^{2}}{\hbar}\sum_{n}\int\frac{d^{2}k_{n}}{(2\pi)^{2}}f^{0}(\varepsilon_{{\bf k}_{n}})\Omega^{n}_{z}({\bf k}_{n})\label{an-hall}\\
&&\alpha_{yx}=\frac{\pi^{2}}{3}\frac{k_{{\rm B}}^{2}T}{e}\frac{\partial \sigma_{yx}(\varepsilon)}{\partial\varepsilon}\Big\vert_{\varepsilon=\mu}\label{an-ner}
\end{eqnarray}
where ${\bf \Omega}^{n}=-{\rm Im}\left[\langle\nabla_{\bf k}u_{n, \bf k}\vert\times\vert\nabla_{\bf k}u_{n, \bf k}\rangle\right]$ is the Berry curvature with Bloch eigenstates $\vert u_{n, \bf k}\rangle$. The first term, Eq. (\ref{an-hall}), is the intrinsic anomalous Hall conductivity, while the second term, Eq. (\ref{an-ner}), gives the anomalous Nernst conductivity which is the transverse electric current in response to a longitudinal temperature gradient in the absence of a magnetic field.
\par
The formalism introduced here is general for any multiband anisotropic material and all of the thermoelectric properties described by $\mathcal{L}_{n}^{\beta}$ can be found by calculating the longitudinal conductivity which depends crucially on the relaxation mechanisms while the intrinsic transverse transport is determined only by the Berry curvature and band structure. Note that this semiclassical theory is valid only in the regime
of dilute impurity concentration {\it i.e.}, when the density of impurities is much smaller than the density of charge carriers.

\section{Numerical Results and Discussion}\label{sec:results}

Here, we present our numerical and analytical results based on the aforementioned theory. We focus on the charge conductivity, Seebeck coefficient ($\mathcal{S}$) and its corresponding figure of merit $\mathcal{ZT}$, in the presence of both the SR and LR potentials. According to the so-called Thomson relation $\Pi=T\mathcal{S}$, where $\Pi$ is the Peltier coefficient, originated from the symmetry properties of the response matrix elements demanded by Onsager reciprocity, there is a direct relation between Seebeck coefficient and the Peltier coefficient. In the following, we set $T=20$ K in all calculated quantities. Furthermore, we will use $n_i =5.2\times10^{12}$ cm$^{-2}$ for the impurity concentration of both short-range and long-range scatterers. This guarantees that the diluteness criteria will be satisfied for a wide range of chemical
potentials in the following results. This impurity concentration corresponds to the chemical potential approximately $\mu\approx10^{-4}$ eV.

\begin{figure}
\includegraphics[width=8.5cm]{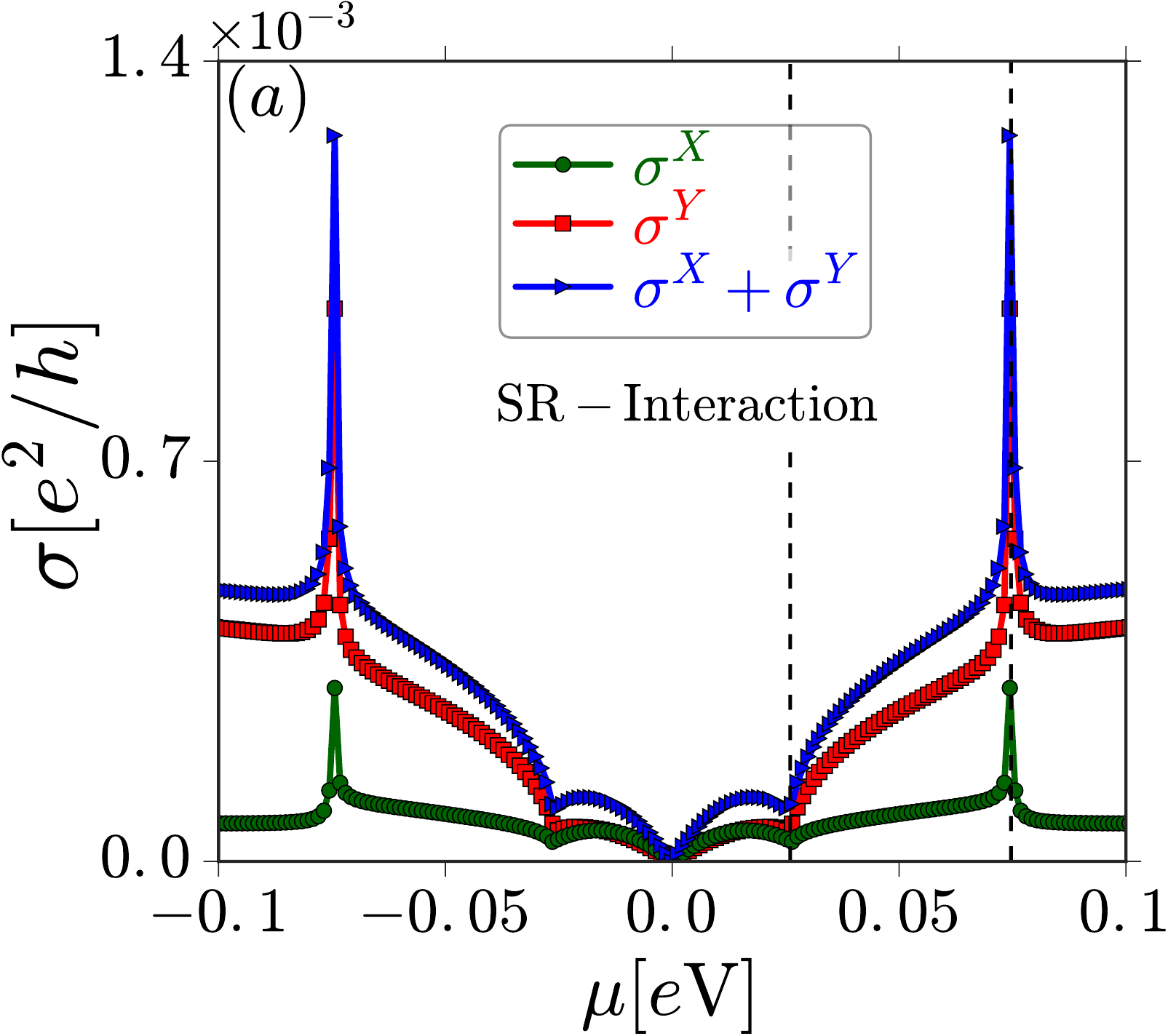}
\includegraphics[width=8.5cm]{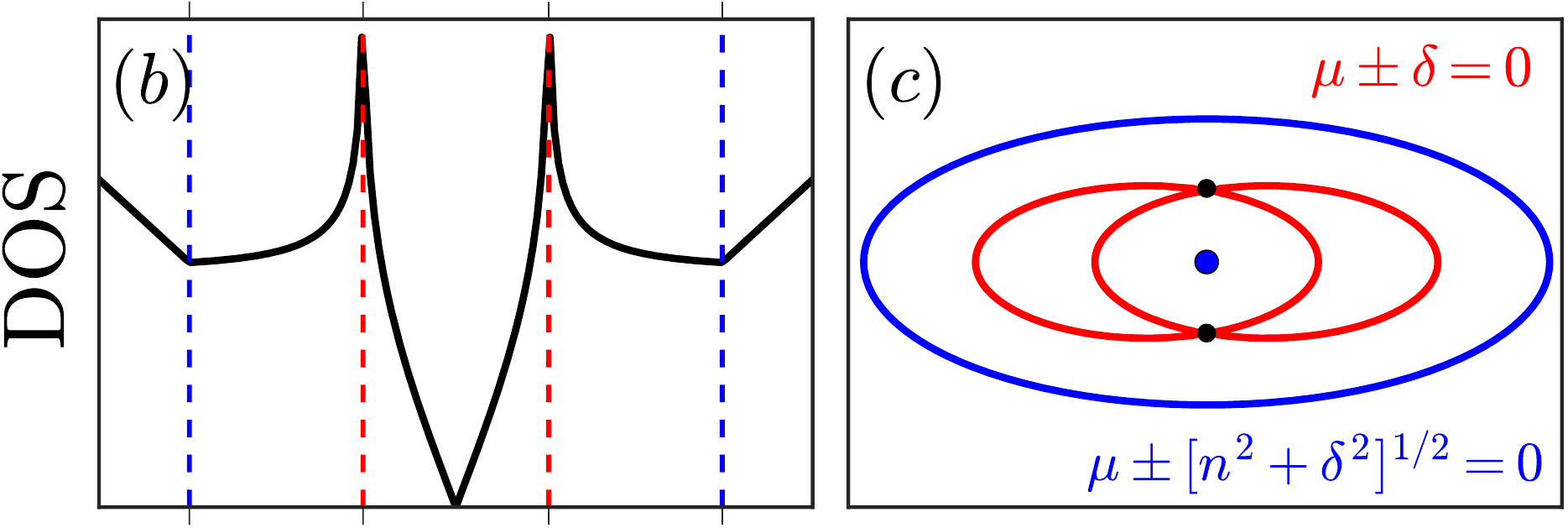}
\caption{(Color online) (a) The $xx-$component of the conductivity of TCIs (001) as a function of doping $\mu$ around the $X$ and $Y$ points at the presence of short-range impurity interaction. There is a local minimum at transition point and a maximum at high-energy Dirac point in the conductivity. (b) Density of states as a function of chemical potential. At transition point, a saddle point on the constant-energy contour leads to a Van Hove singularity at energy $\mu\pm\delta=0$, indicated by red dashed lines. (c) The constant-energy contour when the Fermi energy lies at transition point, $\mu\pm\delta=0$, and high-energy Dirac points, $\mu\pm\sqrt{n^{2}+\delta^{2}}=0$, saddle points marked by black dots and blue dot denotes the high-energy Dirac point.}\label{fig2}
\end{figure}
\par
Figure (\ref{fig2}) shows the $xx$-component of the conductivity around both the $X$ and $Y$ points and the total conductivity as a function of the chemical potential $\mu$, in the presence of SR impurity interaction where we consider $V({\bf k}-{\bf k'})=V_0=1000$ eV{\AA}$^{2}$. Note that this value has been used for graphene \cite{sarma} and we assume that it should be applicable here. The total conductivity $\sigma=\sigma^{X}+\sigma^{Y}$, is simply superpositions of the conductivities contributed by states around the $X$ point and those around the $Y$ point, each one explicitly breaks the corresponding fourfold rotational symmetry. Let us first focus on the conductivity around the $X$ point, due to the $C_{4}$ symmetry the same analysis can be explored for that around the $Y$ point too. The $yy$-component of the conductivity, based on this symmetry is the same as its $xx$-component. The conductivity shows a local minimum at the transition point $\mu=\pm\delta$ originated from the saddle points in the surface states band structure which lead to a Van Hove singularity in the density of states. This can be understood from the fact that the relaxation-rate-like quantities vary inversely with the density of states, so the transition point corresponds to a local minimum in these quantities and as a result a local minimum in the conductivity occurs.

Remarkably, in TCIs materials, there is a maximum at high-energy Dirac points $\mu=\pm\sqrt{n^{2}+\delta^{2}}$ since at these points the density of states for one of the subband vanishes and subsequently no inter-band scattering can happen. This leads to an increasing in the relaxation-rate-like quantities and a subsequently the conductivity increases. The maximum in the conductivity arises from the fact that the states around the high-energy Dirac point, with energies $\varepsilon\approx\pm\sqrt{n^{2}+\delta^{2}}$, which effectively contribute in transport, are relatively better conducting than other states since interband scattering is suppressed to them. A maximum in the conductivity around the Dirac point, due to spin-flip-induced transitions of electrons between exchange split spin subbands, in magnetic graphene has been investigated \cite{zare2013}. It is important to mention that a large electrical conductivity is usually found in high carrier concentration metals.
\begin{figure}
\includegraphics[width=8.5cm]{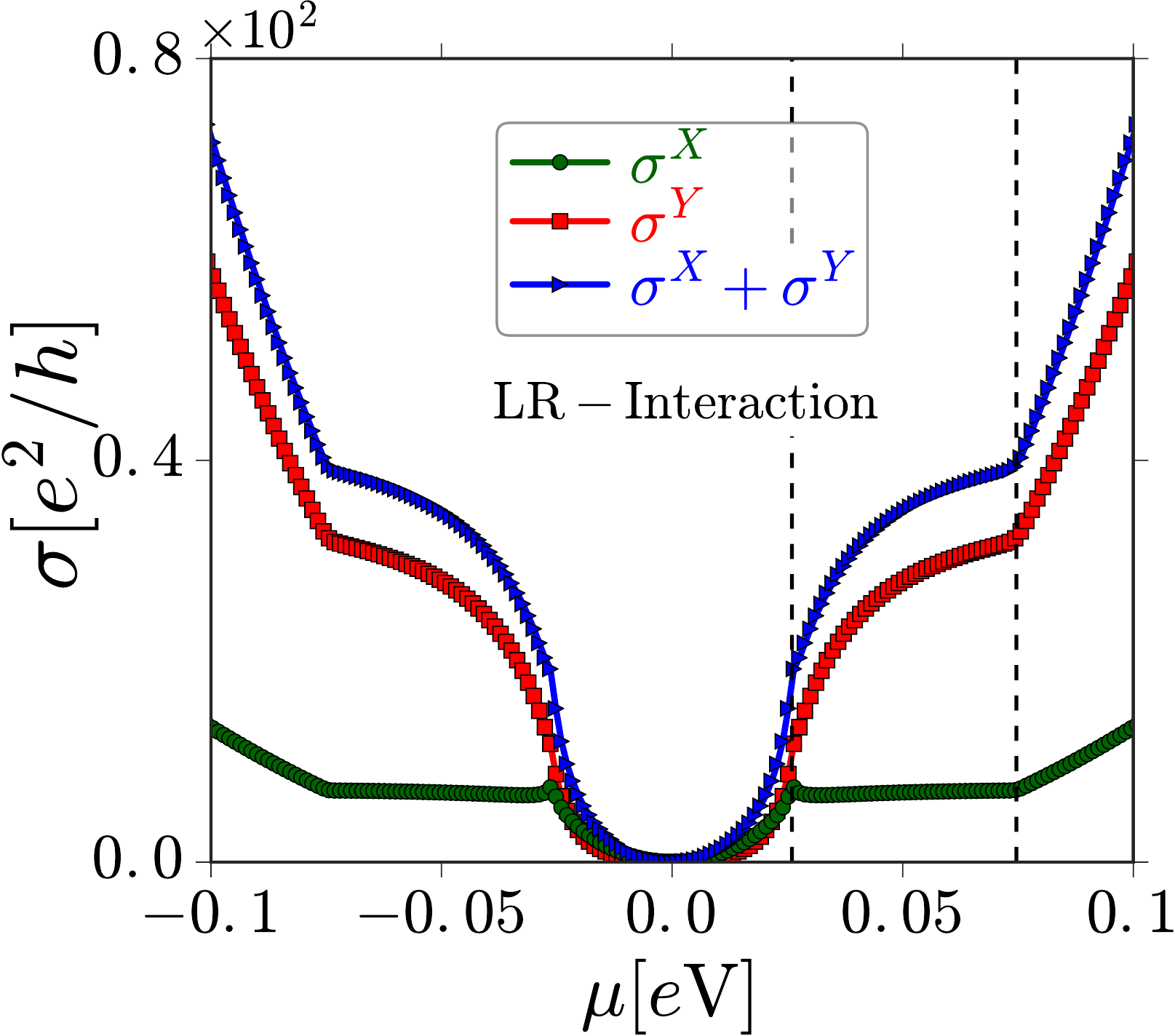}
\caption{(Color online) The $xx-$component of the conductivity of TCIs (001) as a function of doping $\mu$ around the $X$ and $Y$ points at the presence of long-range charge-impurity Coulomb interaction. The dashed lines are guides for the eye.}\label{fig3}
\end{figure}
\par
Owing to the fact that long-range charge-impurity Coulomb interactions are the dominant scatterers in most samples, we thus consider the Coulomb interaction. To account
partially for screening and to avoid the well-known Fermi velocity
artifact of mean-field theory in systems with long-range
interactions~\cite{vignale}, we have used an interaction potential including
Thomas-Fermi screening

\begin{equation}
v_{{\bf k}-{\bf k^\prime}}=\frac{2\pi e^2}{\varepsilon_{0}(|{\bf k}-{\bf k^\prime}|+q_{_{TF}})}~,
\end{equation}
where $q_{_{TF}}=2\pi e^2 N(\mu)/\varepsilon_{0}$ is the Thomas-Fermi screening
vector and $N(\mu)$ is the density of states of the system which is calculated numerically and shown in Fig~2b. The dielectric constant of the insulator crystal SnTe~\cite{dielectric} is quite large and it is about $\varepsilon_{0}=1200$.

In Fig. (\ref{fig3}), the $xx-$component of the conductivity of TCIs (001) as a function of doping $\mu$ around the $X$ and $Y$ points are shown in the presence of LR interactions. Although the energy dependence of the conductivity is quantitatively different in comparison with SR impurities, interestingly the overall behavior for $|\mu|>\delta$, illustrating of exterma, is the same as SR interactions. In other words, our results show that despite the details of scattering phenomena, the band structure and dispersion of TCI surface states play a main role in the conductivity in the high energy regime. On the there hand, there is a clear discrepancy between two type of scatters in the low-energy regime. It should be noted that unlike the graphene~\cite{sarma, zare2015} in which diffusive transport caused by SR impurities do not lead to any thermoelectric effects, due to the constant conductivity, the conductivity of TCI surface states when only short-range scatterers are present has an explicit energy dependence. Furthermore, LR interactions plays essential role in low-energy, however the SR scatter illustrates the most impact at high-energy. Notice that the value of the conductivity is proportional to the value of $V_0^{-2}$ or $\varepsilon^2$ in the SR and LR interactions, respectively and those values can be reduced by considering the surface charge screening effects. It should also be noted that as we concentrate on low temperatures where only electrons contribute effectively in thermal transport, the thermal conductivity behaves more or less like charge conductivity at very low temperatures, originated from Sommerfeld expansion.
\par
\begin{figure}
\includegraphics[width=8.5cm]{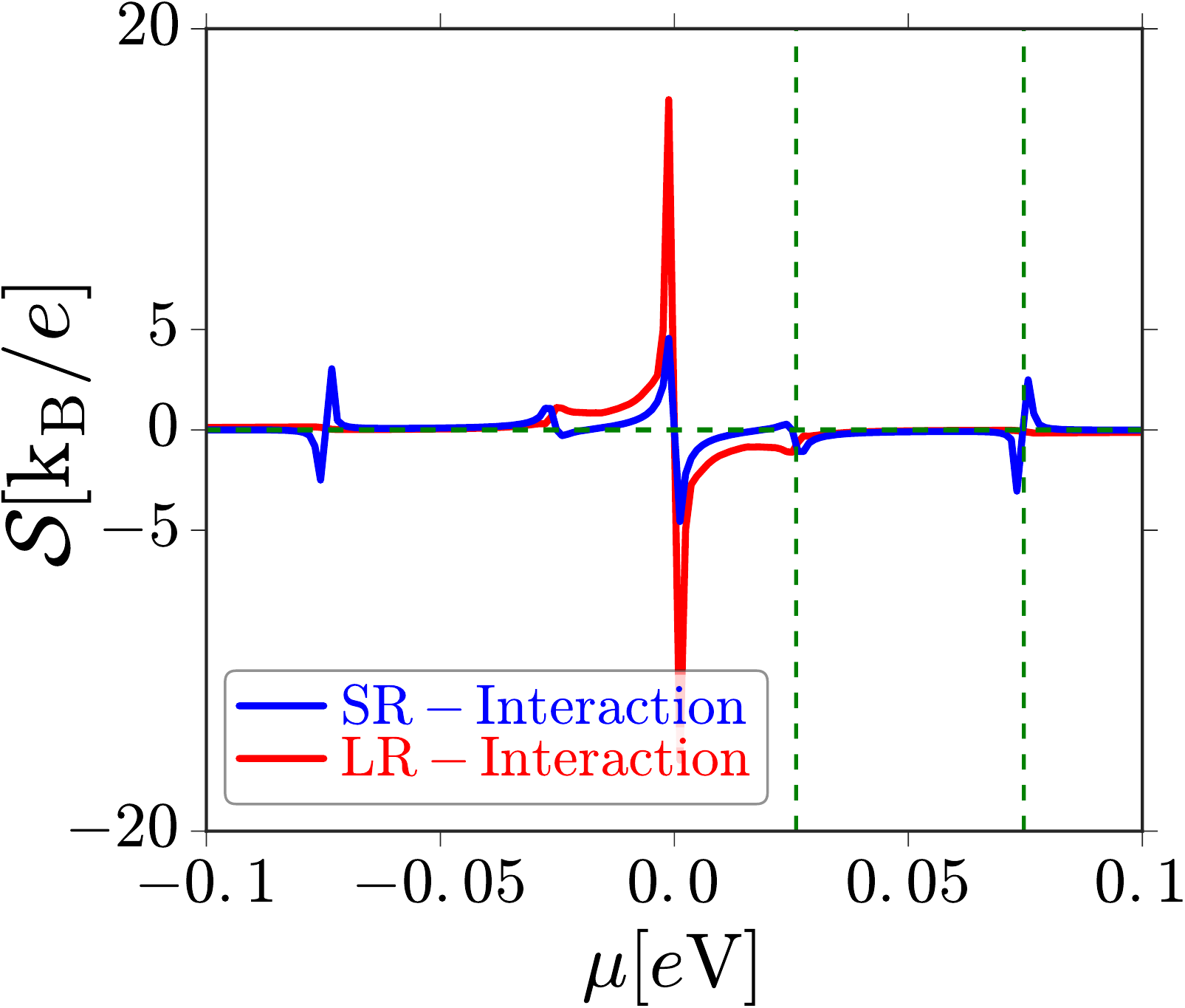}
\caption{(Color online) Seebeck coefficient of TCI (001) surface states as a function of doping $\mu$ at the presence of short- and long-range Coulomb impurities. Vertical dashed lines represent $\mu=\delta$, and high-energy $\mu=\sqrt{n^{2}+\delta^{2}}$, respectively.}\label{fig4}
\end{figure}

Now, we turn to the discussion on the Seebeck coefficient (thermopower) ${\cal S}$ which is more feasible quantities in real experiments. According to Eq. (\ref{lls}), $\mathcal{L}_{n}^{(1)}(\theta, \theta^{\prime})$, which plays the key role in thermoelectric effects, vanishes when $\sigma_{n}(\varepsilon; \theta, \theta^{\prime})$ is a symmetric function of $\varepsilon_{n}-\mu$, owing to the fact that the electron-hole asymmetry around the Fermi level in the band structure or transport properties is responsible for the thermoelectric effects. The variation of thermopower ${\cal S}$ with doping at the presence of SR and LR interactions is obtained as shown in Fig. (\ref{fig4}). We see that the coefficient ${\cal S}$ passes from zero and changes sign when the Fermi level lies at the low- and high-energy Dirac points $\mu=0$ and $\mu\pm\sqrt{n^{2}+\delta^{2}}=0$, respectively, as well as at the transition points $\mu\pm\delta=0$. This is similar to the well-known effect in semiconductors in which the thermopower for $n$ and $p$ types has opposite sign and based on this effect, devices made of $p-n$ junctions are used for electronic cooling. It is worth mentioning that a large Seeback coefficient is usually found in low carrier concentration semiconductors as it shown in this figure.

\begin{figure}
\includegraphics[width=8.5cm]{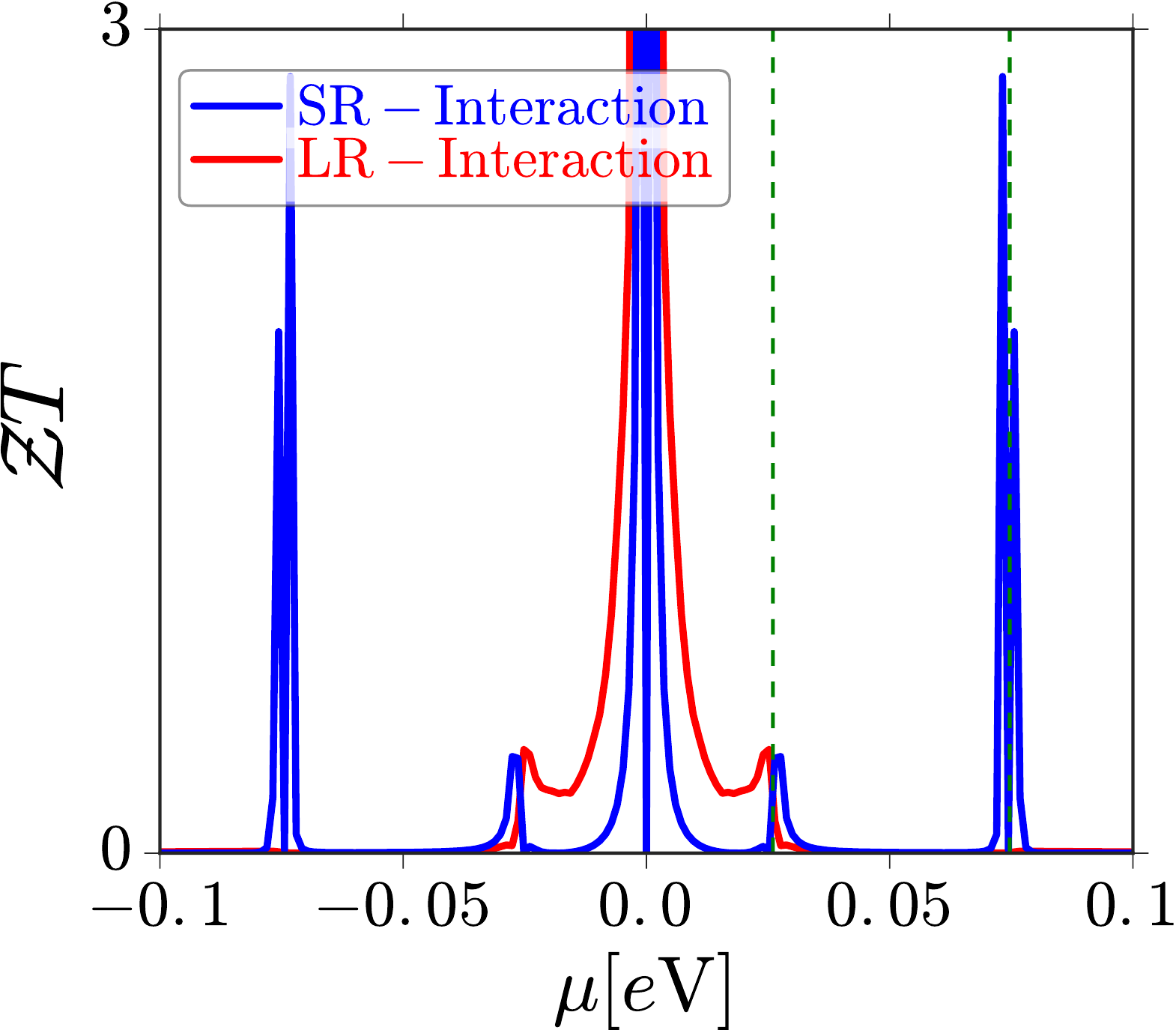}
\caption{(Color online) The variation of corresponding figures of merit are depicted as a function of the chemical potential $\mu$ at the presence of short- and long-range charge-impurity interactions. Vertical dashed lines represent $\mu=\delta$, and high-energy $\mu=\sqrt{n^{2}+\delta^{2}}$, respectively.
}\label{fig5}
\end{figure}
\par
When the Fermi level lies in the region $\vert\varepsilon\vert<\delta$ thermally activated electrons (holes) move along a temperature gradient which result in a charge accumulation gradient in the opposite (same) direction due to the negative (positive) charge of the electrons (holes). Therefore a negative (positive) thermopower is obtained. Notice that in this case, the Fermi surface consists of two disconnected Dirac pockets centered away from the $X$ and $Y$ points of the same type of carriers as indicated in Fig. \ref{fig1}. By further increasing in the energy, then the constant-energy contour evolves rapidly and undergoes a change in topology through a Lifshitz transition, changing from two disconnected Dirac pockets to two concentric pockets of different carrier types. In the conduction band the Fermi surface comprises of a large electron and a small hole pockets both centered at the $X$ ($Y$) point, while in the valance band it consists of a large hole and a small electron Dirac pockets. When the Fermi energy lies in this region, different charge carriers are thermally activated simultaneously. Such excitations carry a different charge current and as a result, their contributions in the Seebeck effect compete to each other. With increasing energy away from $\varepsilon=\pm\delta$ (saddle points) the large pocket becomes even larger, while the small pocket shrinks and eventually vanishes at the high-energy Dirac points, $\varepsilon=\pm\sqrt{n^{2}+\delta^{2}}$, which means the density of states of one subband tends to zero. Consequently, when the Fermi level passes the high-energy Dirac points, once again ${\cal S}$ changes sign. At higher energies, $\vert\varepsilon\vert>\sqrt{n^{2}+\delta^{2}}$, the Fermi surface consists of two concentric Dirac pockets of the same type carriers both centered at $X$($Y$)-point. We see that thermopower always shows the change of the sign in the vicinity of the low- and high-energy Dirac points as well as at the transition point. In addition, as expected, the figure of merit attains its maximum value around the chemical potential $\mu$. The figure of merit of the Seebeck effect becomes large where the thermoelectric effect is very strong while the heat transport is not. This effect can be seen in Fig. \ref{fig5} where the variation of ${\cal Z}T$ is shown with the chemical potential $\mu$. 

\begin{figure}
\includegraphics[width=8.5cm]{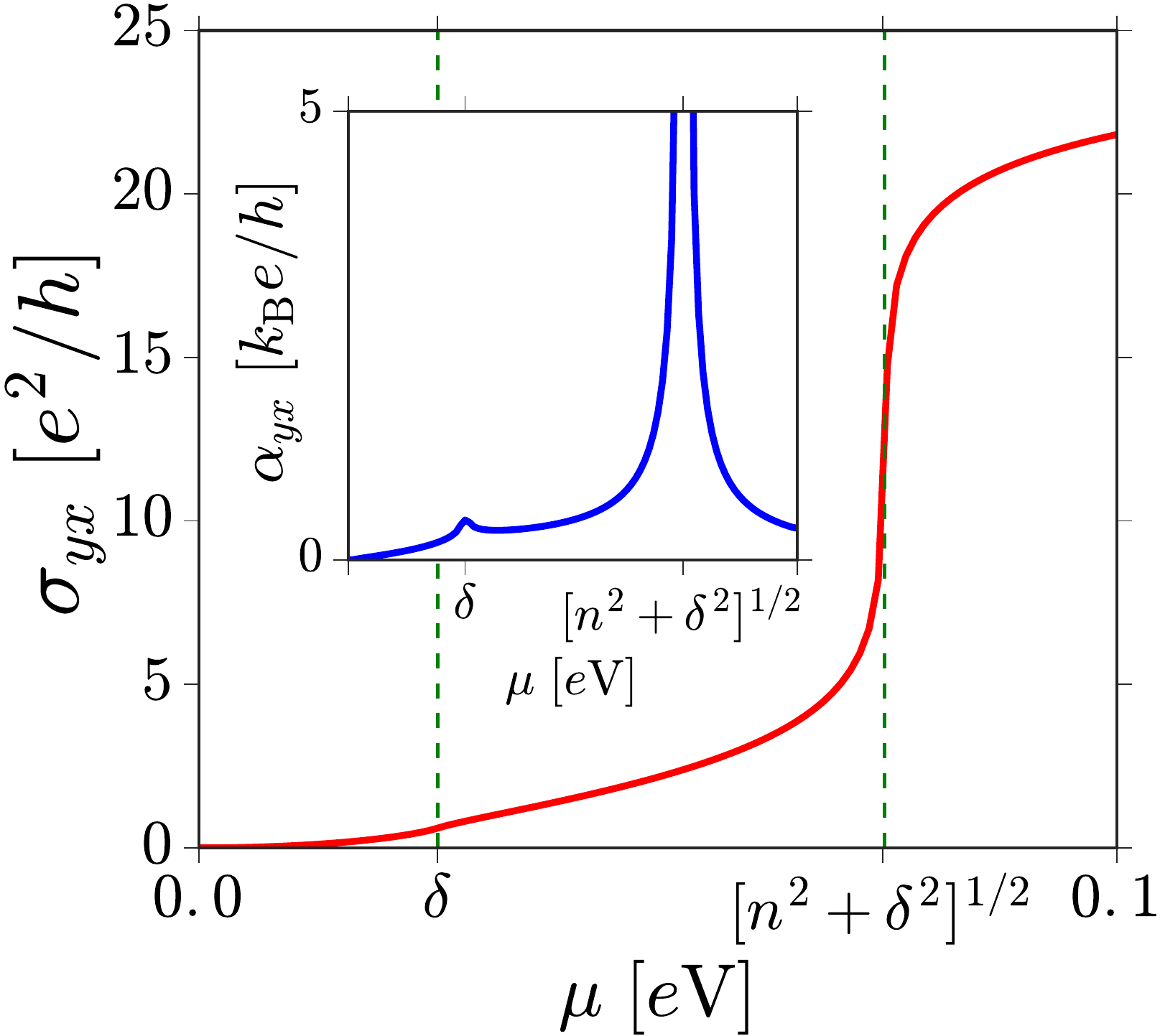}
\caption{(Color online) The variation of anomalous Hall conductivity $\sigma_{yx}$ as a function of the chemical potential $\mu$. In the inset: Anomalous Nernst conductivity $\alpha_{yx}$ as a function of the chemical potential. Vertical dashed lines represent $\mu=\delta$, and the high-energy $\mu=\sqrt{n^{2}+\delta^{2}}$, respectively.}\label{fig6}
\end{figure}

The calculated anomalous Hall conductivity $\sigma_{yx}$ and anomalous Nernst conductivity $\alpha_{yx}$ (inset figure) are plotted in Fig. (\ref{fig6}) as a function of doping. The anomalous Hall conductivity increases almost monotonically with doping below the high energy Dirac point at $\mu=\sqrt{n^{2}+\delta^{2}}$ but enhances significantly when the high energy band also contribute in the transport. This gives rise to a large peak in the corresponding intrinsic Nernst conductivity at high energy Dirac point beside the small peak at transition point $\mu=\delta$.
\par
It is worthwhile comparing the thermoelectric properties of TCIs with other advanced two-dimensional crystalline. Although graphene has very high charge mobility, it has two major problems with a view to thermoelectric applications~\cite{graphene}. First, graphene is a gapless semimetal and would be difficult to separate the opposite contributions of electrons and holes to the Seebeck coefficient. The second difficulty lies in its very high lattice conductivity leads to the decreasing of the $\mathcal{Z}T$. Monolayers of dichalcogenides, on the other hand, have some advantages over gapless graphene and are suitable for many electrical and photonic applications. A famous element in this group is monolayer MoS$_2$ which is a two-valley direct gap semiconductor. It has been shown that the thermal conductivity of MoS$_2$ is three orders of magnitude lower than that of graphene~\cite{liu} and it is insensitive to width and edge-type in a MoS$_2$ nanoribbon. Based on the density-functional theory, it has been investigated~\cite{huang} that $\mathcal{Z}T>0.5$ for the $p$-type monolayer MoS$_2$. Moreover, Fei {\it et al}~\cite{fei} predicted that $\mathcal{Z}T$ can reach to unity at room temperature in phosphorene, a new atomistically thin two-dimensional materials, at low density.
\par
Finally, we should like to comment on the possible influences of phonon in our results which we have not considered. The main effect of phonon is their contribution in the thermal conductivity ${\cal K}$ and the thermopower are not affected with the presence of phonon. Therefore, it is clear that the thermal conductivity of phonon ${\cal K}_{\rm ph}$ can only affect the figure of merit in our results and since it does not depend on the chemical potential, it might only increase ${\cal K}$ depending on the temperature values. This should decrease ${\cal Z}T$ but dependence on the $\mu$ will not be changed qualitatively. At higher temperatures, phonon becomes important but as we mentioned before it only results in the overall decline of the predicted figures of merit, without affecting their qualitative behavior.

\section{Conclusion}\label{sec:concl}
In conclusion, we have studied the effects of both short-range and long-range charge-impurity interactions on the electronic transport and thermoelectric effects of a surface states of topological crystalline insulators using the generalized semiclassical Boltzmann theory for multiband anisotropic systems. Taking into account both intra- and interband scattering processes, we have obtained that the doping dependence of the conductivity exhibits a local minimum at a transition point, originated from the Van Hove singularity in the density of states. The conductivity also illustrates a maximum owing to inter-band-induced transitions of electrons between subbands. This effect is the direct result of the gapless Dirac spectrum of TCIs in which the density of states declines linearly with varying the energy toward the Dirac point with a vanishing the density of states. We also reveal that TCIs could be very promising material for caloritronics~\cite{bauer} studies and applications. Thermopower changes sign several times due to the conversion of electrons to holes and vice versa at each extremum of the conductivity. The intrinsic anomalous transport due to the Berry curvature is also investigated which will be more important in the case of short-range impurities. Therefore, based on this study, we believe that TCIs can be used as a base material to investigate thermoelectric phenomena.

\acknowledgments
We thank F. Parhizgar and F. Ghamsari for fruitful discussions. This work was partially supported by Iran Science Elites Federation.

\end{document}